# Medium-Grain Niobium SRF Cavity Production Technology for Science Frontiers and Accelerator Applications

G. Myneni[1,2,3], Hani E. Elsayed-Ali[2], Md Obidul Islam[2], Md Nizam Sayeed[2], G. Ciovati[3,5], P. Dhakal[3], R. A. Rimmer[3], M. Carl[4], A. Fajardo[4], N. Lannoy[4], B. Khanal[5], T. Dohmae[6], A. Kumar[6], T. Saeki[6], K. Umemori[6], M. Yamanaka[6], S. Michizono[6], and A. Yamamoto[6,7]

[1]BSCE Systems Inc., Yorktown, VA 23693
[2]Electrical and Computer Engineering Department, Old Dominion University, Norfolk, VA 23529
[3]Thomas Jefferson National Accelerator Facility, Newport News, VA 23606
[4]ATI Specialty Alloys & Components, Albany, OR 97321
[5]Department of Physics, Old Dominion University, Norfolk, VA 23529
[6]High Energy Accelerator Research Organization (KEK), Tsukuba, Ibaraki, Japan, 305-0801
[7]European Organization for Nuclear Research (CERN), Geneva, Switzerland, CH-1211

## Executive Summary

We propose cost-effective production of medium grain (MG) niobium (Nb) discs directly sliced from forged and annealed billet. This production method provides clean surface conditions and reliable mechanical characteristics with sub-millimeter average grain size resulting in stable SRF cavity production. We propose to apply this material to particle accelerator applications in the science and industrial frontiers. The science applications require high field gradients (>~40 MV/m) particularly in pulse mode. The industrial applications require high $Q_0$ values with moderate gradients (~30 MV/m) in CW mode operation. This report describes the MG Nb disc production recently demonstrated and discusses future prospects for application in advanced particle accelerators in the science and industrial frontiers.

## Introduction

World's science frontier programs and SRF accelerator applications demand high performance and cost-effective SRF accelerator technology [1-8]. Fine-grain (FG) and Large-grain (LG) niobium (Nb) SRF cavity production have matured and been applied in many present-day accelerator projects and research programs. We have recently developed medium-grain (MG) Nb material and demonstrated cost-effective Nb disc production, by directly slicing discs from a forged and annealed Nb billet.

The established FG Nb sheet production process is very complex, requiring more than 15 steps making the sheets prone to contamination. It is very expensive to produce FG Nb because of stringent QA control requirements. The LG Nb disc is produced by directly slicing the as-cast ingot. It is relatively simple and straight forward to keep the surface clean. Thus the LG Nb disc production cost is significantly lower compared to the FG Nb sheet production. However, there are drawbacks concerning the non-uniform mechanical characteristics which can be attributed in part to the random grain orientation of each large grain. This results in broad variation of mechanical properties and shape related issues during the elliptical cavity forming process. Regardless, the LG cavities have achieved the high-gradient and excellent $Q_o$ performance goals with lower cost compared to FG Nb sheet.

The MG Nb disc production has been recently demonstrated. The MG Nb discs are directly sliced from a forged and annealed billet. The process requires fewer steps than the FG Nb sheet, resulting in a reduced production cost [9]. These discs are expected to be superior to LG Nb as they tend to be more homogeneous and mechanically stable with the sub-millimeter average grain size and acceptable mechanical characteristics.

The features of the MG Nb production path are compared with the FG and LG production paths in Fig. 1. The production of FG Nb sheet requires 15 steps, including forging (6. forging) and rolling (8~10). Conversely, the production of LG Nb material requires 7 fewer steps compared to the FG Nb sheet. The reduction in processing steps leads to a more cost effective material and avoids the risk of contamination during the forging and rolling processes. However, the LG Nb material has grains much larger than 1 cm that lead to a broad distribution of mechanical characteristics within a single disc. This can cause issues in SRF cavity fabrication. Additionally, the averaged tensile strength may be relatively lower than FG and MG Nb material.

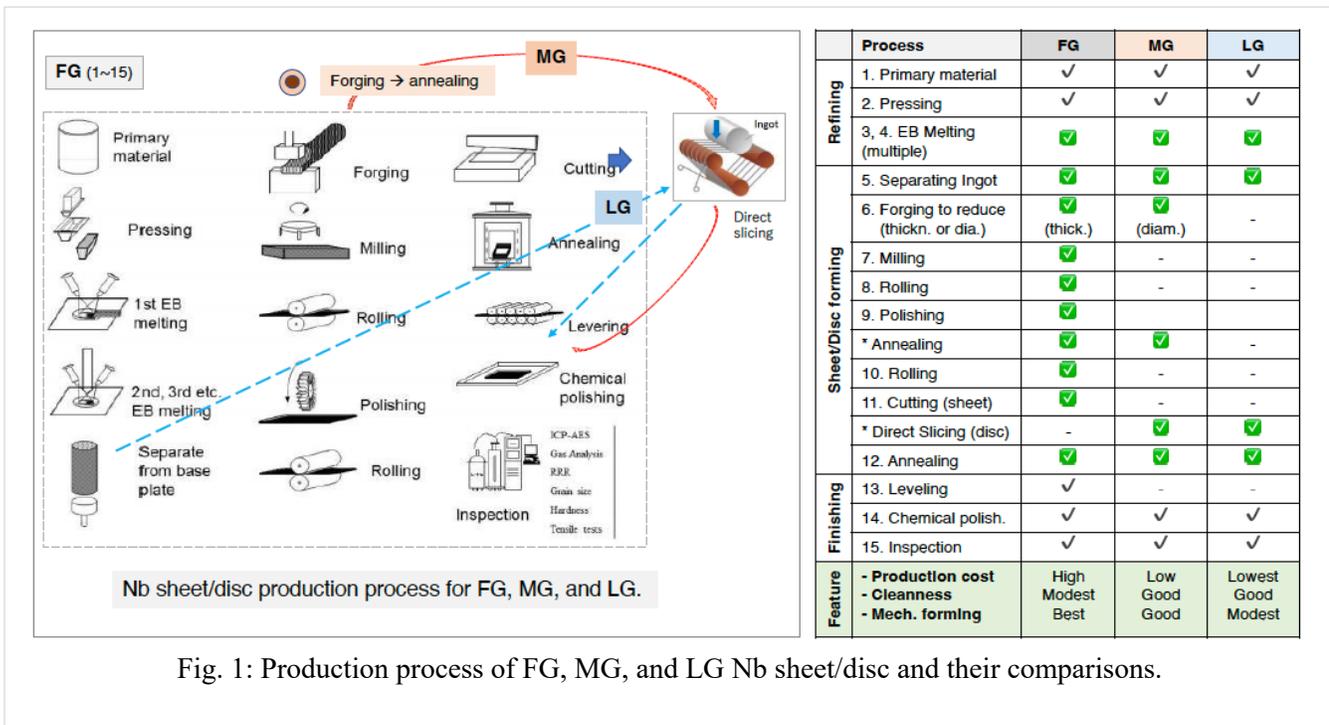

Fig. 1: Production process of FG, MG, and LG Nb sheet/disc and their comparisons.

The production of the MG Nb discs eliminates 5 processes, as shown in Fig. 1. This forging and annealing process produces an average grain size <1mm with occasional grains as large as 5mm. The mechanical properties are more uniform than LG Nb and are closer to the FG

Nb sheet. The smaller grain-size in the forged and annealed billet provides a narrower distribution of mechanical properties than that of the as-cast LG Nb and therefore certification of the MG Nb material for the HPGS regulation becomes practical, as discussed later. Moreover, the process flow of the MG Nb material importantly eliminates the rolling process which reduces the risk of contamination. Even though adding two more processes compared to the LG Nb material, the MG Nb material is still cost-effective compared to FG Nb sheet production.

We propose further development of the MG Nb SRF cavity production path for particle accelerator applications in the science and industrial frontiers.

## MG Niobium Development and their Physical Characteristics

In this section the production of Nb discs from forged and annealed billets is discussed, the resulting mechanical properties and RF performance are also discussed. Fig. 2 shows the direct slicing process in the production of LG and MG Nb discs. LG Nb discs are produced by directly slicing the as-cast ingot on a wire saw. The MG Nb discs are sliced from forged and annealed billet.The resulting microstructure of each process is shown on the right hand side of Fig. 2, the LG Nb has several >>1cm grains while the MG Nb has an average grain size of 0.2-0.3mm with occasional grains as large as 5mm. The forged and annealed billet was provided by ATI to KEK and BSCE/JLab who had the discs sliced from the billet based on their prior experience. The SRF cavities have been fabricated and RF tested by KEK and JLab to qualify this approach.

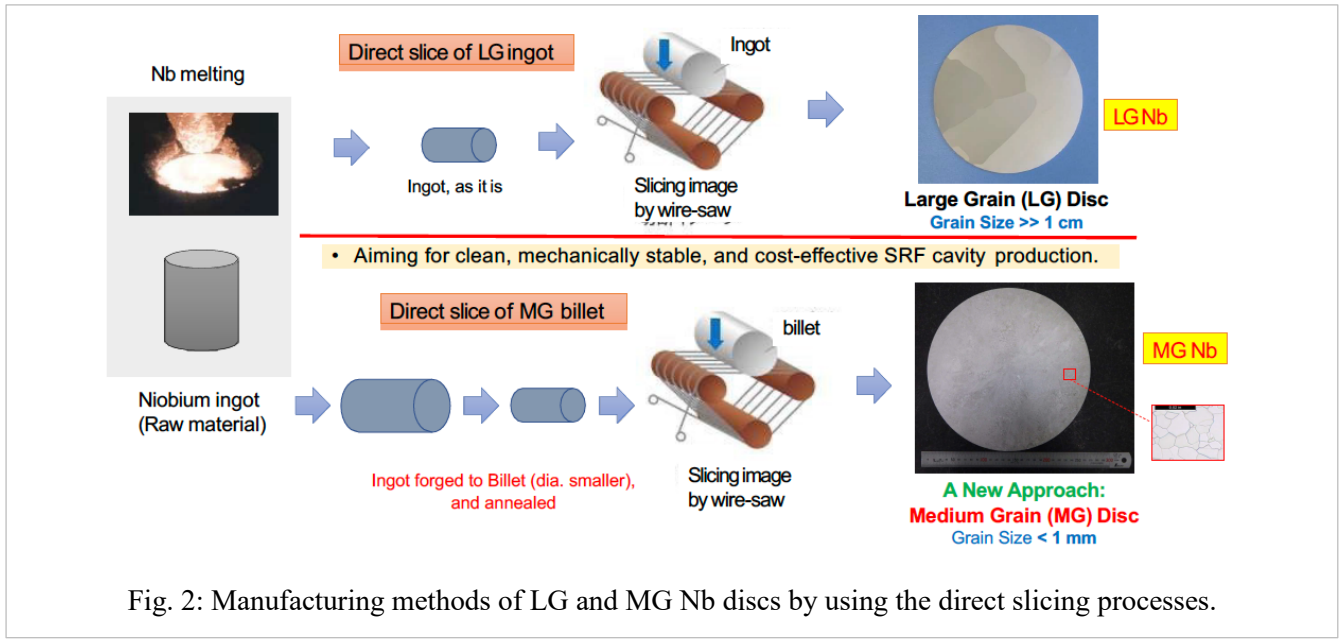

Fig. 2: Manufacturing methods of LG and MG Nb discs by using the direct slicing processes.

Table 1 provides a summary of the MG Nb material properties. The purity of the MG Nb is high as the measured RRR values of the MG Nb billet range from 450 to 523. The attached cross sectional photo shows the average grain-size is around 0.2 ~ 0.3 mm and is fully recrystallized. The measured value of yield-strength (0.2%) at room temperature is 61 MPa and this is comparable to the typical value of FG Nb material ($\geq$ 50 MPa). Also, the measured values of tensile-strength at room temperature are 141 and146 MPa and are comparable to the typical value of FG Nb material (≥ 140 MPa) [10].

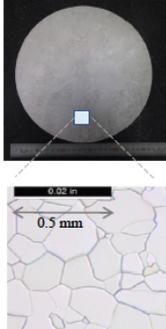

Table 1. Material properties of MG Nb material produced by ATI.

Chemical composition (typ.):
Ta: < 0.01%
W: < 0.003%
Ti, Si, Mo, Fe: < 0.003%
Ni: < 0.002 %
$H_2$ < 0.0003%
C: < 0.002%
$N_2$: < 0.002%
$O_2$: < 0.004%

Dimensions: 260 m (dia.) 2.8 mm (t)
RRR: > 450
Recrystallization: 100%
Grain size: 0.2 ~ 0.3 mm (ASTM: 1~2)
Hardness (HV0.1): 40 ~ 44
Mechanical Strength:
  Ultimate strength (RT) > 141, 146 N/mm²
  Yield strength (RT) > 56, 61 N/mm²

Comparisons of mechanical properties of the FG and MG materials measured at room temperature are shown in Fig. 3 [11,12]. The red horizontal lines in the plot denote the criteria of the HPGS regulation for the KEK/STF 9-cell cavity: the required yield strength is 39 MPa, the required tensile strength is 120 MPa. The measured average values of yield-strength and tensile-strength exceed the criterion (the red line) for both FG and MG Nb materials. The measured values of elongation are also shown in Fig. 3. The elongation of the MG Nb material is 24 % and is lower than that of FG Nb (37%). The lower elongation of MG Nb material is not an issue for certification under the HPGS regulation. However, the lower elongation (<30%) at room temperature needs to be improved for reliable cavity fabrication. During the elliptical press forming process a low percent elongation increases the risk of cracks forming at the iris edge of the half-cell cup. Some cracks around the iris can be repaired; however, this additional work and the potential loss of a half cell reduce the cost-effectiveness and total cavity yield of this SRF cavity production method.

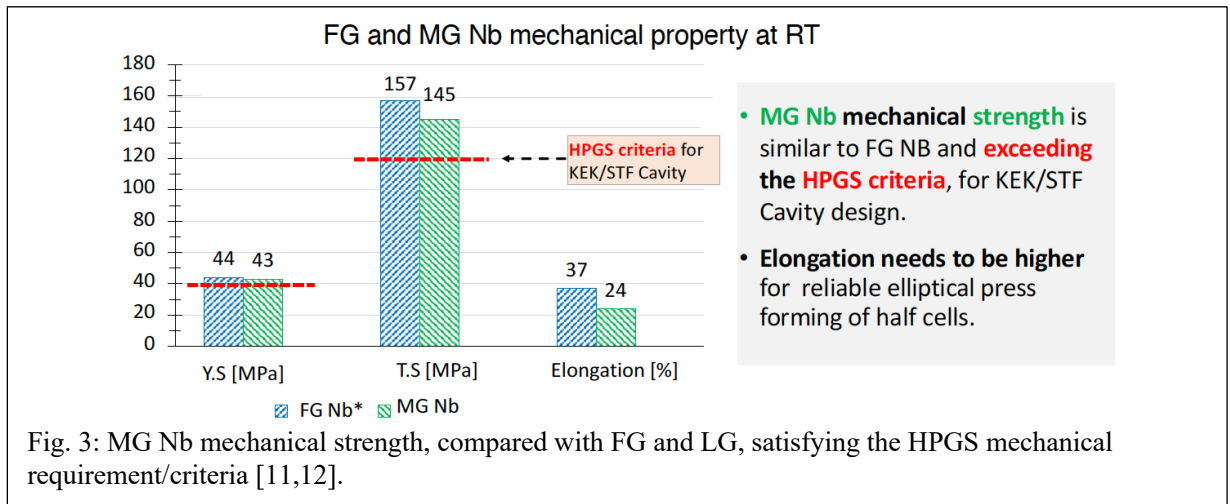

Fig. 3: MG Nb mechanical strength, compared with FG and LG, satisfying the HPGS mechanical requirement/criteria [11,12].

Two 1.3GHz single-cell cavities were fabricated from the MG Nb discs at KEK. The two cavities were treated by the standard surface preparation (bulk-EP, annealing at 800 C for 4 hours, the final EP, and 120 C baking for 48 hours). The measured SRF performances of these two SRF single-cell cavities (R18 and R18b) are shown in Fig. 4. The accelerating gradients of both cavities have exceeded the ILC specification of 35 MV/m. The results are also compared with the performance of FG single-cell cavity (R9). The SRF performances of single-cell cavities made from FG and MG Nb material are comparable. Details of SRF performance tests are found in the reference [13].

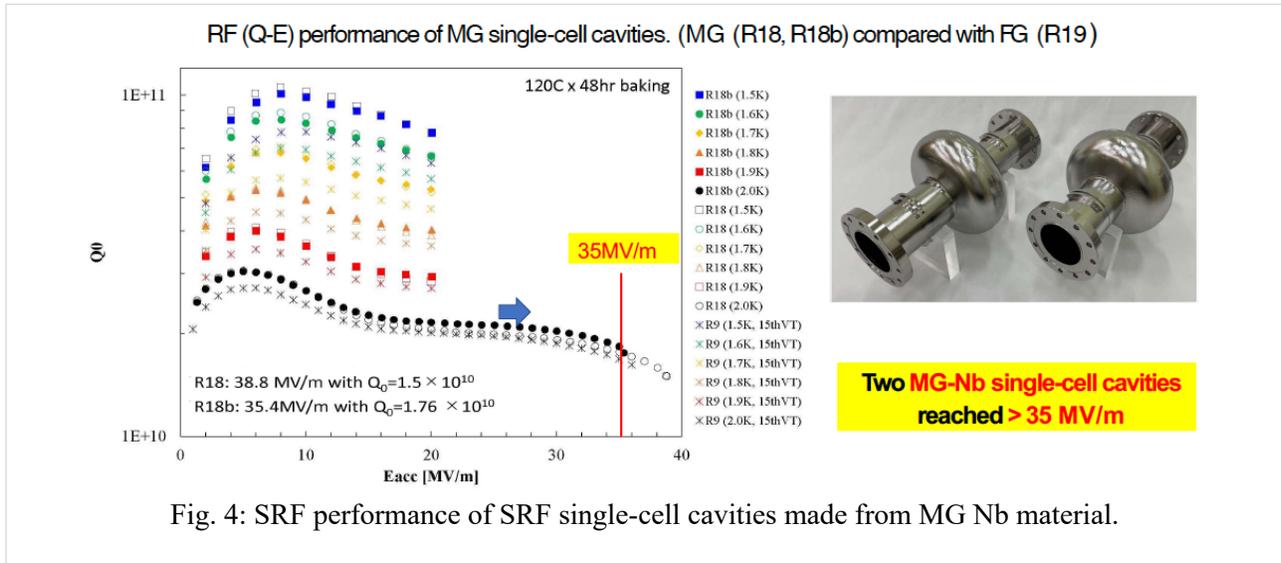
Fig. 4: SRF performance of SRF single-cell cavities made from MG Nb material.

Based on the successful achievement of the MG Nb material and the SRF performance demonstrating > 35 MV/m, further development of the MG Nb material is in progress as follows:

- Reproducibility of the MG Nb material performance to be tested with a second billet manufactured by ATI,
- Improvement of mechanical properties, specifically elongation, to facilitate more reliable cavity fabrication, and
- Trial for 9-cell cavity production to verify the RF performance to be achieved.

In summary, MG Nb disc has been successfully manufactured and the feasibility for future applications have been demonstrated. The MG Nb discs display mechanical characteristics and RF performance comparable to FG Nb sheet. MG Nb is expected to contribute to the particle accelerator applications in the science and industrial frontiers, as discussed below.

## MG Nb Cavity Applications for Science Frontiers

The MG Nb disc directly sliced from the billet may provide promising features of (i) clean surface, (ii) sufficiently strong and homogeneous mechanical properties and (iii) adequate formability, allowing cost-efffective cavity production, as described above. MG Nb may be applied to future SRF accelerators in the science and industry frontiers. We focus here on the applications for the International Linear Collider (ILC) program as an anticipated future program for a Higgs factory.

ILC is an energy-frontier electron-positron collider based on two key technologies: (i) superconducting radio-frequency (SRF) and (ii) nano-beam technologies. The center-of-mass energy (C.E.) will be in a range of 200–500 GeV, extendable to 1 TeV [14-16], with a unique feature of the linear accelerator with the luminosity as high as $10^{34}$ cm$^{-2}$.s$^{-1}$, based on the SRF technology. Based on the discovery of the Higgs boson with a mass of 125 GeV at CERN, the plan has been updated to start as the "Higgs factory" with a C.E. 250 GeV and with the total length of the accelerator complex to be is 21 km [15,16]. The project is still to be authotrized with a global frame work.

The accelerator system is composed of i) a polarized electron and positron sources, ii) damping rings (DR) at 5 GeV, iii) beam transport followed by a two-stage bunch-compressors accelerating the beam up to 15 GeV, iv) 5.5 km main linacs accelerating the beam up to 125 GeV by using SRF cavities with an average gradient of 31.5 MV/m, and nanoscale focused beam-delivery systems which bring the beams into collision with a 14 mrad crossing angle at a single interaction point. The ILC accelerator and SRF cavity parameters are summarized, and the ILC accelerator configuration and layout is shown in Fig. 5.

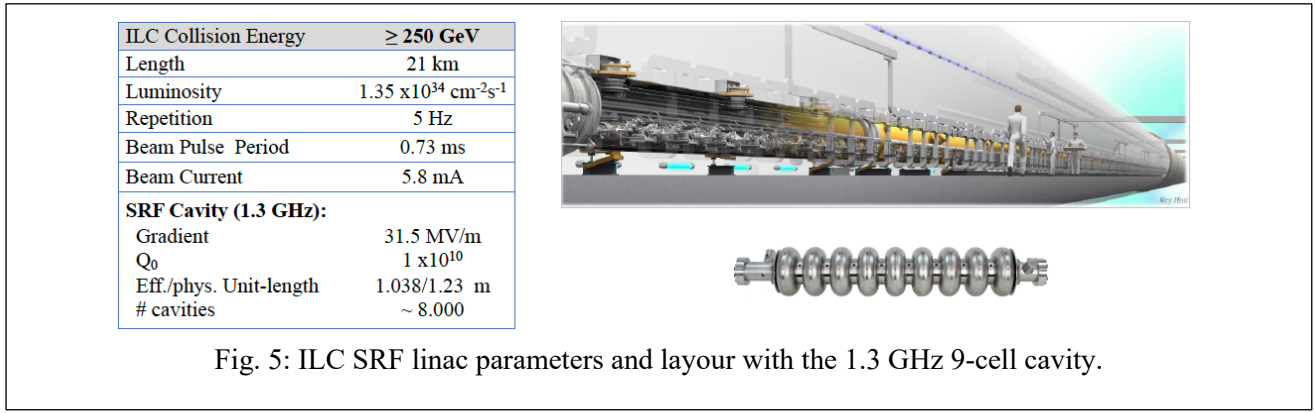

Fig. 5: ILC SRF linac parameters and layour with the 1.3 GHz 9-cell cavity.

The cost of niobium materials for fabrication of SRF cavities is relatively high, owing to the use of a rare material and an elaborate preparation process. R&D aims to reduce the cost of materials by optimizing the production of Nb ingots and by optimizing the disc/sheet and the elliptical cavity forming process, to prepare for cavity fabrication. The TDR (and also XFEL, LCLS-II) specified the residual resistivity ratio (RRR) to be >300 [17]. Low RRR material may constrain achievable maximum cavity highest gradients, thus we target to optimize the purity of ingots with a lower RRR $\geq$ 300, and anticipate to simplify the Nb disc production using direct slicing from MG Nb billet, to maintain clean surfaces. We expect a major cost reduction of the fabrication of Nb sheets while keeping the superior cavity RF performance. The technical requirement proposal for the Nb sheet/disc for the ILC SRF cavity is summarized in Table 2.

Table 2. Technical requirements proposal for Nb discs/sheets from the ILC 1.3 GHz SRF cavity:

| Chemical composition : | RRR $\geq$ 300 |
|---|---|
| Ta : $\leq$ 0.05%  (< 0.15%, tbd) |     - *RRR value supercedes chemistry  (tbd)* |
| W : $\leq$ 0,007% | Grain size (FG and MG to be accepted): |
| Ti : $\leq$ 0.005% |     - *to be relaxed from: < 50 μm (on average, FG), to  < 1 mm (on average, MG)* |
| Si : $\leq$ 0.005% | |
| Mo : $\leq$ 0.005% | Mechanical Strength: |
| Fe : $\leq$ 0.003 % |     - Ultimate strength (RT): > 120 N/mm$^2$ |
| Ni : $\leq$ 0.003 % |     - Yield strength (RT): > 40 N/mm$^2$ |
| |     - Elongation (RT):  > 30 % (expected) |
| H$_2 \leq$ 2 : weight ppm (tbd) |     - Hardness (Hv10: > ~ 40 (tbc) |
| N$_2$, O$_2$, C:  10 weight ppm (tbd) | Dimensions:  260 mm (dia.)x  2.8 mm (t) for central 9-cell parts. |
| |     - # Sheets/Discs ~ 2 x 9 x 8,000 x 1.1 = ~ 160,000  (~ 210 tons), |

The MG Nb material may be applicable to the ILC SRF cavity production, by updating the grain size requirement to < 1 mm (on average, ASTM 00) from the current requirement of <~0.05 mm (ASTM 4~5) with assuming the FG Nb sheet. We may accept FG and MG Nb sheet/disc with satisfying RRR and mechanical property requirements. On the other hand, we may need to reserve the possibility of LG depending on the improvement of the mechanical strength and the stability. We aim to enlarge the Nb material availability/market in cooperation with worldwide Nb material suppliers to co-work on the large amount of Nb sheets/discs production (within a time period of several years with optimized work sharing). Careful study and negotiation will be required to get authorization from the HPGS regulation, and we believe that it can be accomplished.

## MG Nb Cavities for Industrial Applications

The MG Nb cavity development motivated by the industrial applications has been carried out as a subset of the CEBAF cryomodule refurbishment program (C75) at Jefferson Lab [18]. Two single cell cavities by using ATI MG Nb of RRR~ 100 have been fabricated with conventional methods of deep drawing of the Nb discs and electron beam welding of the half-cell and beam tubes. The shape deviation of the half-cells was inspected with a 3D laser scanner and ~63% of the points were within ±0.1 mm from the ideal shape as shown in Fig. 6. This value is consistent with what was achieved with a standard FG Nb disc, using the same dies, and it is better than achieved with LG Nb [19].

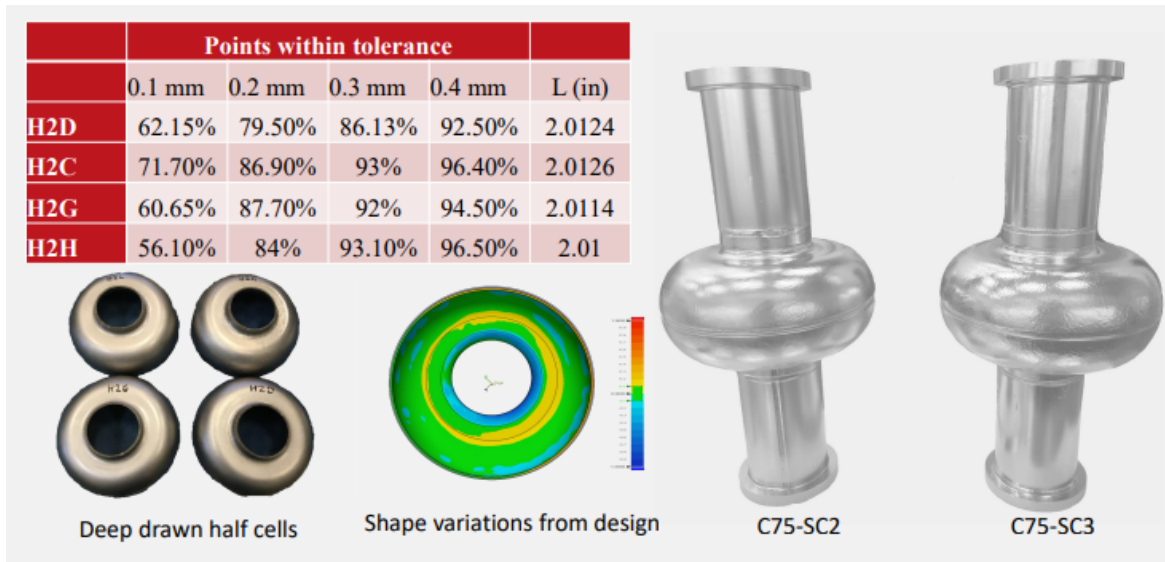

Fig. 6: Fabrication of tow single cell cavities labeled C75-SC2 and C75-SC3.

After the fabrication with electron beam welding, the cavities received ~120 μm of inner surface removal by electropolishing (EP) followed by the vacuum annealing at 650 °C for 10 hours. The cavities again received 20 μm inner surface EP. Standard procedures were followed to clean the cavity surface in preparation for an RF test: degreasing in ultrapure water with a detergent and ultrasonic agitation, high pressure rinsing with ultrapure water, drying in the ISO 4/5 cleanroom, assembly of flanges with RF feedthroughs and pump out ports and evacuation. The cavity was inserted in a vertical cryostat and cooled to 4.2 K with liquid helium using the standard

Jefferson Lab cooldown procedure in a residual magnetic field of < 2 mG. The cooldown of the cavities were done such that the residual flux trapping during the normal to superconducting state is minimum. The RF measurements were done at 2.0 K acquiring $Q_0(E_{acc})$.

Fig. 7 shows the RF performance of the two single cell cavities measured at 2.0 K. The cavity C75-SC2 was limited at 14 MV/m with quench and C75-SC3 reached 30 MV/m with $Q_0$ ~1.25×10$^{10}$ limited by quench. Both cavities were again subjected to 120 °C bake for 48 hours in ultra-high vacuum and the RF test at 2 K was repeated. The performance of C75-SC2 did not change as a result of 120 °C bake, however, C75-SC3 reached 35 MV/m with $Q_0$ ~ 9.2 ×10$^9$. The cavity encountered multipacting ~ 25 MV/m, but was easily processed with high RF power. Multipacting is an unfortunate feature of the cell shape. Both cavities were tested with oscillating superleak transducers mounted on the test stand and the cavities' quench location was identified close to cavity equator. An optical inspection of the inner surface at the quench location showed an overall roughness and a possible weld defect at the equator of C75-SC2.

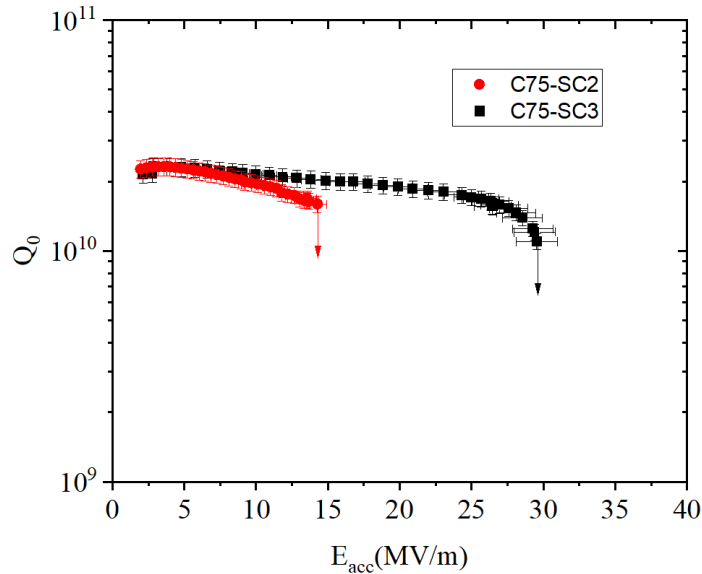

Fig. 7: The RF performance of two single cell cavities at 2.0 K after 650 °C/10 h followed by 20 μm EP.

To further explore the RF performances, both cavities were subjected to a 2-step centrifugal barrel polishing (CBP) with medium and fine polishing media, removing ~ 60 μm from the inner surface followed by 30 μm EP. The cavities received heat treatment at 800 °C/3 h with Nb caps installed in the beam tube to protect the cavities inner surface from furnace contamination. This step also eliminated the need of post heat treatment EP. RF results showed no significant improvement over the previous RF performances. Fig. 8 shows the RF performance of C75-SC3 after each surface treatments. Fig. 9 shows the images from the optical inspection of the quench region in cavity C75-SC2 before and after the CBP with medium polishing media.

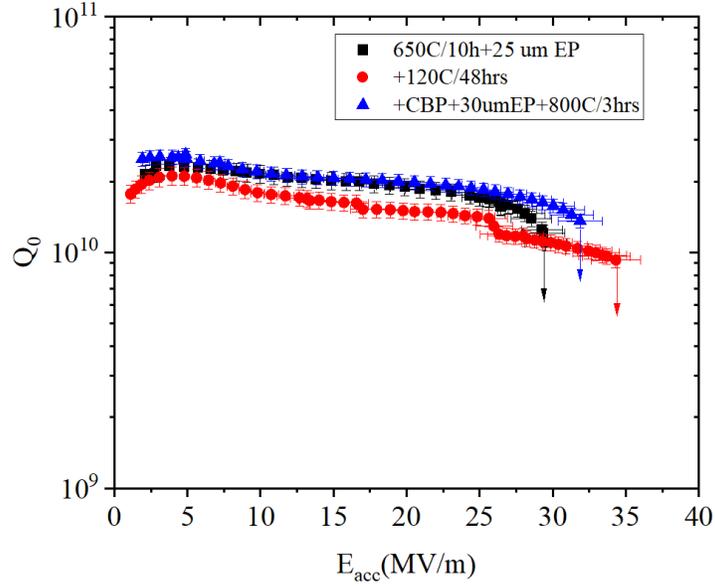

Fig. 8: RF performance at 2.0 K of single cell cavity C75-SC3 after each surface preparations.

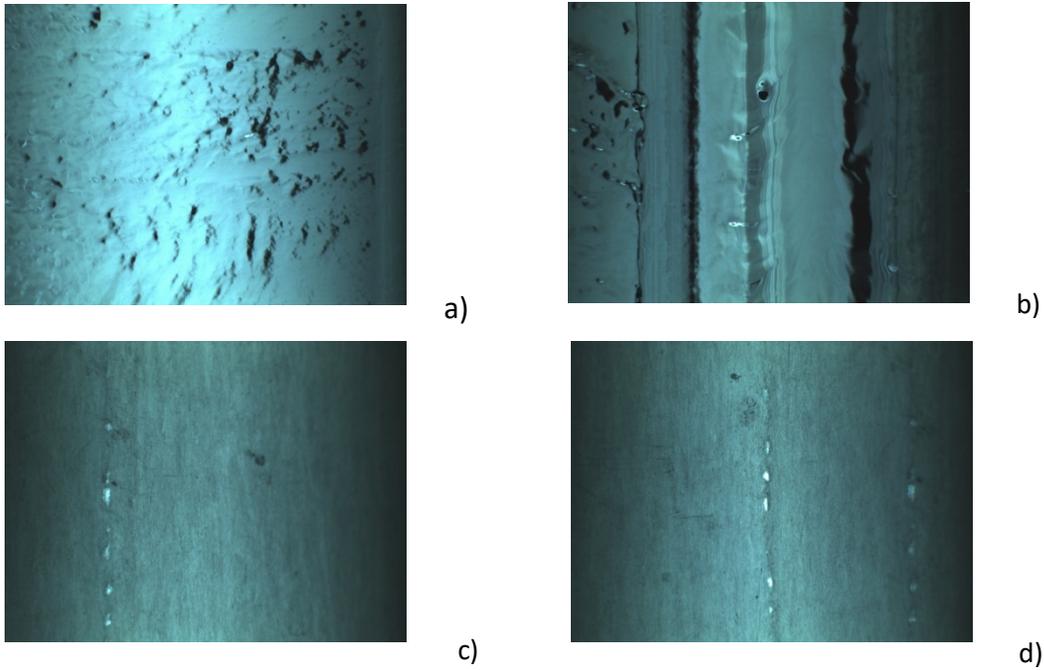

Fig. 9: Optical images of the inner surface of C75-SC2 near the quench location before (a), (b), after CBP (c), (d).

To further evaluate the loss mechanism due to residual magnetic flux trapping during the cavity cooldown, we have measured the flux expulsion. Three flux gate magnetometer were placed on equator of the cavity and the ratio magnetic flux measured during normal to superconducting transition was measured as a function of temperature gradient across the cavity (iris-iris) as shown in Fig. 10. The flux expulsion behavior is poorer compared to cavities made from LG Nb [20] but comparable to cavities made from FG Nb subjected to 800 ºC heat treatments [21]. The cavities have been annealed at higher temperature (1000 ºC/3hrs) and the measurement will be carried out in future.

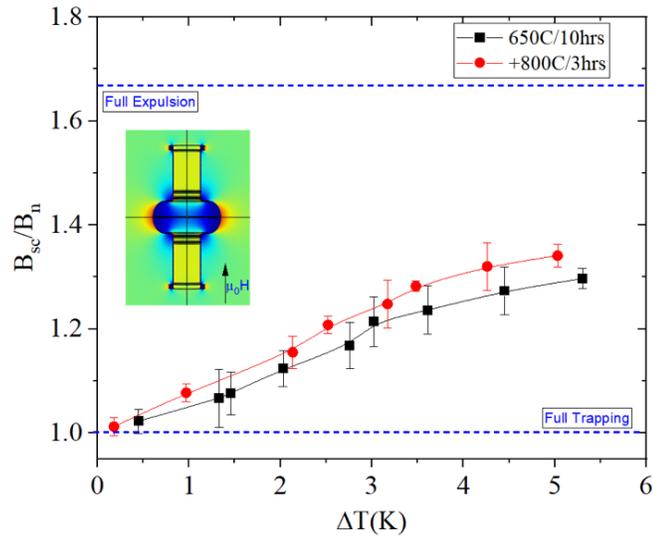

**Fig. 10**: The flux expulsion behavior of C75-SC3 after each heat treatment.

ATI MG Nb discs with RRR~100 were provided by BSCE, and were shipped to Research Instruments GmbH, Germany, to fabricate a 5-cell 1497 MHz cavity. The cavity will be processed with the same treatment procedures currently followed for all other C75 cavities, made of LG Nb, which consists mainly in CBP, EP and vacuum annealing at 800 °C. Fig. 11 shows a set of pictures of one half-cell after deep-drawing, showing a surface texture commonly referred to as "orange peeling" to describe the roughness resulting from the deep-drawing of a material with grain size of a few hundreds of microns.

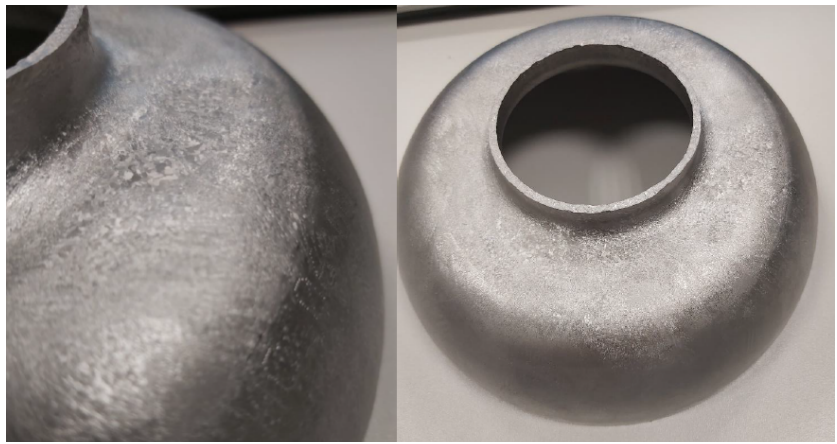

Fig. 11: Picture of a cavity with RRR~100 MG Nb after deep drawing (Courtesy of Research Instruments, GmbH).

If the cavity performance will meet the C75 project specifications ($E_{acc}$ = 19.1 MV/m with $Q_0$ of $8\times10^9$ at 2.07 K), it will be installed in one of the C75 cryomodule which will be likely be installed in CEBAF in 2024. It will demonstrate a model work with MG Nb with RRR ~100, as a very cost-effective SRF cavity toward promising industrial applications.

## Fundamental Study of MG Nb – Thermal Modulation

High accelerator gradients (>40 MV/m) and enhanced $Q_0$ SRF cavities are essential for science frontiers where as high $Q_0$ cavities of modest accelerator gradients (~30 MV/m) are needed for industrial applications. In order to achieve these high performance accelerator structures economically, one needs to understand the negative role played by the subsurface included hydrogen as well as the niobium hydrogen compounds in the grain boundaries [22-25]. The Nb-hydrogen systems' role in the thermal diffusivity of cavity surfaces will be investigated in the ongoing fundamental research discussed here.

Time-domain thermoreflectance (TDTR) is used to measure the thermal diffusivity of Nb and $Nb_3Sn$. In TDRT, an ultrashort pump laser pulse heats the surface region, then a probe pulse monitors the modulation in reflectance with time, which is used to extract the surface temperature. The TDTR setup is schematically shown in Fig. 12 (a). A Ti:sapphire laser (wavelength $\lambda$ = 800 nm, pulse duration $\tau$ ~110 fs, and repetition rate of 80 MHz) is used as the laser source. The laser beam is split into a pump and probe beams. The pump beam is modulated at 600 kHz to allow for phase-locked detection of reflectivity modulation $\Delta R/R$ [26, 27].

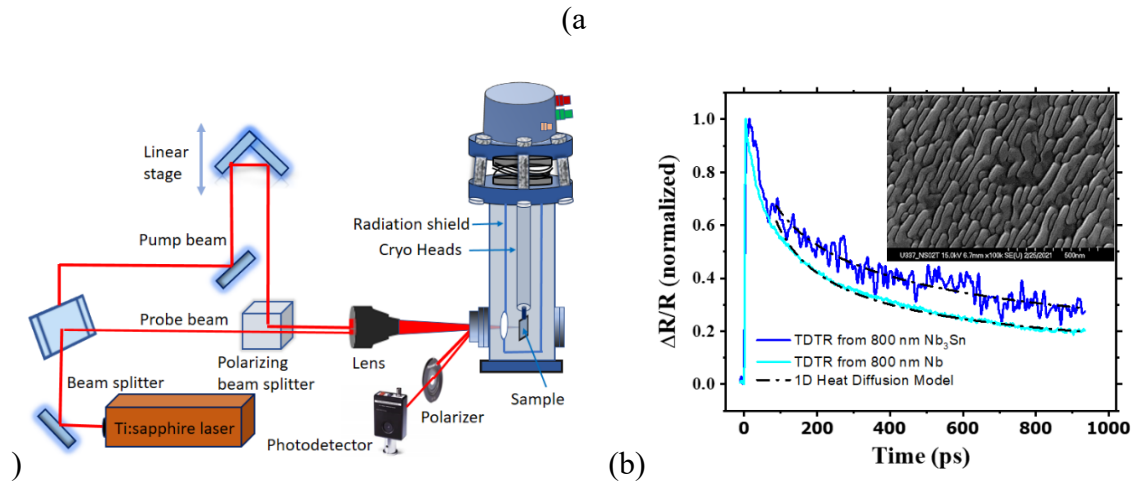

Fig. 12. (a) Femtosecond TDRT pump-probe setup with cryo-cooler. The linear stage is used to generate the time delay between the pump and probe pulses. (b) $\Delta R/R$ of the 800 nm Nb and $Nb_3Sn$ films and 1D heat diffusion model fit to data. The inset is an scanning electron microscopy image of the Nb film showing elongated oriented grains with average width ~50 nm and length ~350 nm.

Niobium on copper and niobium tin ($Nb_3Sn$) on sapphire films were deposited by magnetron sputtering. Fig. 12 (b) shows $\Delta R/R$ from the films taken at room temperature. To determine the thermal diffusivity of the Nb and the $Nb_3Sn$ films, the $\Delta R/R$ scans are fitted to the surface temperature from a one-dimensional (1D) heat diffusion mode based on the parabolic one-step heat transfer model given by [28,29]:

$$\frac{\partial T(x,t)}{\partial t} = -\alpha \frac{\partial^2 T(x,t)}{\partial x^2} + \frac{J(1-R)}{\rho C_p t_p d} * \exp\left[-\frac{x}{d} - 2.77\left(\frac{t}{t_p}\right)^2\right] \quad (1)$$

where, $T(x,t)$ is the temperature profile, $x$ is the distance normal to the surface, $\alpha$ is thermal diffusivity in m$^2$s$^{-1}$, $\rho$ is mass density in kgm$^{-3}$, and $C_p$ is the specific heat capacity in Jkg$^{-1}$K$^{-1}$, $R$ is the reflectivity, $J$ is laser fluence, $d$ is radiation penetration depth, and $t_p$ is pulse width.

From the heat diffusion model fit to the $\Delta R/R$ scans, shown in **Fig. 12 (b)**, $\alpha$ is obtained. For the 800 nm Nb film $\alpha = 23.7\pm0.20\times10^{-6}$ m$^2$s$^{-1}$ which similar to that reported for bulk Nb [30]. For the Nb$_3$Sn film $\alpha = 1.48\pm0.20\times10^{-6}$ m$^2$s$^{-1}$ which is greater than that of the reported value of $1.11\pm0.20\times10^{-6}$ m$^2$s$^{-1}$ measured at room temperature and similar to the diffusivity value measured at 20 K for polycrystalline Nb$_3$Sn [31]. We recently installed a cryogenic cooler to measure $\alpha$ of Nb and Nb$_3$Sn for different processing conditions at temperatures as low as 9 K.

MG Nb and MG Nb$_3$Sn samples will be prepared with various process steps needed for producing high performance cavities and the thermal diffusivity measurements will be carried to optimize these process procedures.

## Conclusions

In summary, the MG Nb disc production has been successfully demonstrated with similar mechanical characteristics and RF performance compared with FG Nb sheet. MG Nb will contribute to future applications in the science and industrial frontiers. However, it will be prudent to optimize the MG SRF cavity process-procedures for obtaining better performance (E$_{acc}$ and Q$_0$) compared to FG cavities. These optimized processes for the cavity fabrication and surface preparation are likely to differ from that of FG cavities. Thus, we propose to extend the science and technology of the MG Nb material as well as the cavity process-procedures.